\documentclass[12pt]{article}
\usepackage[dvips]{graphicx}
\renewcommand{\baselinestretch}{1.20}
\begin{document}

%
\begin{flushright}
OU-HET-582, May, 2007 \ \ \\
\end{flushright}
\vspace{0mm}
\begin{center}
\large{Comparative analysis of the transversities and the longitudinally
polarized distribution functions of the nucleon}
\end{center}
\vspace{0mm}
\begin{center}
M.~Wakamatsu\footnote{Email \ : \ wakamatu@phys.sci.osaka-u.ac.jp}
\end{center}
\vspace{-4mm}
\begin{center}
Department of Physics, Faculty of Science, \\
Osaka University, \\
Toyonaka, Osaka 560-0043, JAPAN
\end{center}

\vspace{4mm}

\vspace{6mm}
\begin{center}
\small{{\bf Abstract}}
\end{center}
\vspace{-2mm}
\begin{center}
\begin{minipage}{15.5cm}
\renewcommand{\baselinestretch}{1.0}
\small
A first empirical extraction of the transversity distributions for
the $u$- and $d$-quarks has been done by Anselmino {\it et al.} based
on the combined global analysis of the measured azimuthal asymmetries in
semi-inclusive deep inelastic scatterings and those in
$e^+ e^- \rightarrow h_1 h_2 X$ processes. Although with large
uncertainties, the determined transversity distributions already
appear to reveal a remarkable qualitative difference with the
corresponding longitudinally polarized distributions. We point out
that this difference contains very important information on
internal spin structure of the nucleon.

\normalsize
\end{minipage}
\end{center}
\renewcommand{\baselinestretch}{2.0}

\vspace{8mm}

As a member of three independent twist-2 parton distribution
functions, the transversity distributions, usually denoted as
$\Delta_T q (x)$, or $h_1^q (x)$, or $\delta q (x)$,
are believed to contain valuable information for our deeper
understanding of internal spin structure of the
nucleon \cite{JJ92},\cite{BDR02}. 
Unfortunately, because of their chiral-odd nature, we cannot access
them directly through the standard inclusive deep-inelastic
scatterings. They can be accessed only through physical processes
which accompany quark helicity flips.
At present, the cleanest way is believed to measure
the transverse spin asymmetry $A_{T T}$ in Drell-Yan processes in
$p \bar{p}$ collisions at high energies
\cite{PAX05}-\nocite{ABDN04}\nocite{EGS04}
\cite{PPB06}.
Another promising (and also practical) way is to measure the so-called 
transverse single-spin asymmetries in the semi-inclusive deep inelastic
scatterings \cite{JLAB07}.
A main drawback here as compared with the Drell-Yan measurement is our 
limited knowledge on the spin-dependent fragmentation mechanism
implemented by the so-called Collins function \cite{Collins93}.
What gave a drastic breakthrough toward the success of this
strategy is the recent independent measurement of the Collins
function in unpolarized $e^+ e^- \rightarrow h_1 h_1 X$ 
processes by the Belle Collaboration at KEK \cite{Belle06}.
Armed with this new information, Anselmino {\it et al.}
carried out a combined global analysis \cite{ABDKMPT07} of the
azimuthal asymmetries in semi-inclusive deep inelastic scatterings
measured by the HERMES \cite{HERMES05} and
COMPASS Collaborations \cite{COMPASS07}, and those in
$e^+ e^- \rightarrow h_1 h_2 X$ processes by the Belle
Collaboration \cite{Belle06}.
Although with large uncertainties, this enables them to determine
the transversity distributions and the Collins functions of the
$u$- and $d$-quarks, simultaneously.
Their main result for the transversities can be summarized as follows.
The transversity distribution is positive for the $u$-quark and negative
for the $d$-quark, the magnitude of $\Delta_T u$ is larger than that of
$\Delta_T d$, while they are both significantly smaller than the
corresponding Soffer bounds \cite{Soffer95}.
From the theoretical viewpoint, the last observation, i.e. the
fact that the transvestites are significantly smaller than the
corresponding Soffer bound seems only natural.
It is because the magnitude of the unpolarized 
distributions are generally expected to be much larger than the
polarized distributions. In our opinion, what is more interesting
from the physical viewpoint is the comparison of the
transversities with the longitudinally polarized 
distributions.

A main purpose of the present study is to perform a comparative analysis 
of the transversities and the longitudinally polarized distribution
functions in light of the new empirical information on the transversities
obtained by Anselmino {\it el al.} \cite{ABDKMPT07}.
We shall show that their results already indicate a remarkable
qualitative difference between these twist-2 spin-dependent distribution 
functions, which in turn contains valuable information for clarifying
internal spin structure of the nucleon.

As is widely known, the most important quantities that characterize the 
transversities are their 1st moments called the tensor charges. They are 
to be compared with the axial charges defined as the 1st moments of the 
longitudinally polarized distributions.
Because of their fundamental importance, they were already investigated 
in various theoretical models \cite{Olness93}\nocite{HJ95}
\nocite{KPG96}\nocite{HJ96}\nocite{SS97}\nocite{GRW98}
\nocite{WK99}\nocite{Wakamatsu01}\nocite{SUPWPG01}\nocite{ETZ04}
-\cite{PPB05} as well as in the
lattice QCD simulations \cite{ADHK97},\cite{Kuramashi98}.
Within the simplest model of baryons, i.e. the nonrelativistic quark model,
no difference appears between the axial and tensor charges.
This means that the difference between the axial and tensor charges is 
purely relativistic effects. As emphasized in \cite{WK99},
however, one must clearly distinguish two types of relativistic
effects. The one is dynamical 
effects, which generates sea-quark polarization. The other is kinematical 
effects, which make a difference between the axial and tensor charges even 
though the sea quark degrees of freedom are totally neglected.
The existence of the latter effect can most easily be seen by remembering
the predictions of the MIT bag model \cite{JJ92},\cite{HJ95},
i.e. a relativistic ``valence quark 
model'' for the isoscalar and isovector axial and tensor charges :
\begin{eqnarray}
 g_A^{(I = 0)} &=& \,1 \cdot \int \,
 \left(f^2 - \frac{1}{3} \,g^2 \right) \,r^2 \,d r, \ \ \ \ \ 
 g_A^{(I = 1)} \ = \ \frac{5}{3} \cdot \int \,
 \left(f^2 - \frac{1}{3} \,g^2 \right) \,r^2 \,d r, \\
 g_T^{(I = 0)} &=& \,1 \cdot \int \,
 \left(f^2 + \frac{1}{3} \,g^2 \right) \,r^2 \,d r, \ \ \ \ \ 
 g_T^{(I = 1)} \ = \ \frac{5}{3} \cdot \int \,
 \left(f^2 + \frac{1}{3} \,g^2 \right) \,r^2 \,d r,
\end{eqnarray}
where $f(r)$ and $g(r)$ are upper and lower components of the lowest energy
quark wave functions.
For a typical bag radius $R \simeq 4.0 \,\omega_1/M_N$ used in
\cite{JJ92}, this gives
\begin{eqnarray}
 g_A^{(I = 0)} &\simeq& 0.64, \ \ \ \ \ g_A^{(I = 1)} 
 \ \simeq \ 1.07, \\
 g_T^{(I = 0)} &\simeq& 0.80, \ \ \ \ \ g_T^{(I = 1)} 
 \ \simeq \ 1.34,
\end{eqnarray}
or equivalently
\begin{eqnarray}
 \Delta u &\equiv& g_A^u \ \simeq \ 0.86, \ \ \ \ \ \ 
 \Delta d \ \equiv \ g_A^d \ \simeq \ -0.21, \\
 \delta u &\equiv& g_T^u \ \simeq \ 1.07, \ \ \ \ \ \ 
 \delta d \ \equiv \ g_T^u \ \simeq \ -0.27.
\end{eqnarray}
This should be compared with the predictions of the CQSM at the model
energy scale around $Q^2 \simeq (600 \,\mbox{MeV})^2$, which includes
not only the kinematical relativistic effects but also the dynamical
effects of nonperturbative vacuum polarization :
\begin{eqnarray}
 g_A^{(I = 0)} &\simeq& 0.35, \ \ \ \ \ g_A^{(I = 1)} \ \simeq \ 1.31, \\
 g_T^{(I = 0)} &\simeq& 0.68, \ \ \ \ \ g_T^{(I = 1)} \ \simeq \ 1.21,
\end{eqnarray}
or equivalently
\begin{eqnarray}
 \Delta u &\equiv& g_A^u \ \simeq \ 0.83, \ \ \ \ \ \ \ 
 \Delta d \ \equiv \ g_A^d \ \simeq \ -0.48, \\
 \delta u &\equiv& g_T^u \ \simeq \ 0.95 \ \ \ \ \ \ \ \ 
 \delta d \ \equiv \ g_T^u \ \simeq \ -0.27.
\end{eqnarray}
One observes that the biggest difference between the predictions of the 
CQSM and the MIT bag model appears in the isosinglet axial charge.
Note that only the prediction of the former model is consistent with
the famous EMC observation, while the latter is not.
In fact, any other effective models
of baryons than the CQSM fail to reproduce such a small value of
$g_A^{(I = 0)}$ around $0.3 \sim 0.4$ \cite{WY91},\cite{WW00}.
(Here, it is assumed to work in the standard $\overline{\mbox{MS}}$
regularization scheme, in which the net longitudinal quark polarization
$\Delta \Sigma$ can be identified with the iso-singlet axial
charge $g_A^{(I=0)}$.)
The isoscalar axial charge is an exception, however.
The other observables are less sensitive to the differences of the models.
For instance, the isoscalar tensor charges predicted by the above two models
are not extremely different as compared with the case of axial charges.

What characteristic features do we expect for the transversities and the 
longitudinally polarized distributions from the above consideration of the 
axial and tensor charges ?
Broadly speaking, we expect that
\begin{eqnarray}
 \Delta q^{(I = 0)} (x) &\ll& \Delta_T q^{(I = 0)} (x), \\
 \Delta q^{(I = 1)} (x) &\simeq& \Delta_T q^{(I = 1)} (x),
\end{eqnarray}
which can alternatively be expressed as
\begin{eqnarray}
 &\,& \Delta u (x) \ > \ 0, \ \ \ \ \ \ \,\delta d (x) \ < \ 0, \\
 &\,& \Delta u_T (x) \ < \ 0, \ \ \ \ \Delta_T d (x) \ < \ 0,
\end{eqnarray}
with 
\begin{equation}
 | \Delta_T d (x) | \ \ll \ | \Delta d (x) | .
\end{equation}

To make the argument more quantitative, we first compare the
CQSM predictions for the transversities and the longitudinally
polarized distributions for the $u$- and $d$-quarks. As for the
longitudinally polarized distributions, we basically use the
results of \cite{WK99} and \cite{Wakamatsu03}, while for the
transversities we use the results obtained in \cite{WK99} and 
\cite{Wakamatsu01}, except one minor modification explained
below. (We recall that, in these studies, the Pauli-Villars
regularization scheme with single-subtraction was used with
the dynamical quark mass of $M = 375 \,\mbox{MeV}$.)
That is, within the framework of the CQSM, the isoscalar
polarized distributions survive only at the 1st order in $\Omega$,
the collective angular velocity of the soliton, which scales as
$1 / N_c$ \cite{WK99},\cite{DPPPW96}-\nocite{DPPPW97}\cite{WGR96}.
On the other hand, the isovector polarized distributions
generally receive contributions not only from the leading
$O(\Omega^0)$ term but also from the subleading $O(\Omega^1)$
term \cite{WK99},\cite{Wakamatsu03}.
The latter subleading correction to $\Delta_T q^{(I=1)}(x)$ was
omitted in the calculation by the Bochum group within the same
model \cite{SUPWPG01}.
However, such $1 / N_c$ corrections are known to be important for
resolving the underestimation problem of the isovector axial charge
$g_A^{(I=1)}$ inherent in the hedgehog soliton models
\cite{WW93},\cite{CBGPPWW94}, so that we included them in
\cite{WK99},\cite{Wakamatsu01}.
Unfortunately, the the vacuum polarization contributions
to $\Delta q^{(I=1)}(x)$ and $\Delta_T q^{(I=1)}$ contained in
this $1 / N_c$ correction term (although they are numerically very
small) turns out to show somewhat peculiar (slowly) oscillating
behavior near $x=0$,
which might indicate some conflict with the basic principle of
relativistic quantum field theory \cite{DPPPW96},\cite{DPPPW97}.
In view of this circumstance, we decided
here to retain only the contribution of ``valence'' level
in this subleading terms of $\Delta q^{(I=1)}(x)$ and
$\Delta_T q^{(I=1)}(x)$, and drop less important Dirac sea
contributions in them. (The terminology ``valence'' here means
quarks in the discrete bound state level coming from the positive
energy continuum under the influence of the hedgehog mean field,
and it should not be confused with the corresponding term in the
parton model discussed shortly.)
To get some feeling about the size of the omitted term, it may be
useful to see its contribution to the isovector tensor charge.
The neglected vacuum polarization contribution to
$g_T^{(I=1)} (\Omega^1)$ is 0.04, which is much smaller than the
corresponding valence quark contribution of 0.36 and the leading
$O (\Omega^0)$ contribution of 0.85 to the same quantity.

In view of the fact that the CQSM reproduces the
phenomenologically known longitudinally polarized distributions quite
well, we think it useful to give its predictions for the transversities
in a simple parameterized form for common use. The fitted transversity
distributions consist of the valence quark part (in the sense of
parton model) and the sea (or antiquark) part as
\begin{equation}
 \Delta_T q(x) \ = \ \Delta_T q_{val}(x) \ + \ \Delta_T \bar{q}(x).
\end{equation}
It turns out that the valence quark parts of distributions are well
fitted in the form :
\begin{equation}
 \Delta_T q_{val} (x) \ = \ a \,\left[ \,
 1 + b \,x + (c \,x^2 + d \,x^3 + e \,x^4) \,e^{- f \,x} \,\right] \,
 (1 - x)^g,
\end{equation}
with
\begin{eqnarray}
 a &=& 0.915395, \ \ \ b \ = \ 2.93304, \ \ \ c \ = \ 129.508, \ \ \ 
 d \ = \ - 361.82, \nonumber \\
 e &=& 271.256, \ \ \ f \ = \ 0.231887, \ \ \ g \ = 2.65858,
\end{eqnarray}
for the $u$-quark, and with
\begin{eqnarray}
 a &=& -0.857512, \ \ \ b \ = \ 12.9987, \ \ \ c \ = \ 32.6664, \ \ \ 
 d \ = \ - 114.033 \nonumber \\
 e &=& 115.414, \ \ \ f \ = \ -5.89189, \ \ \ g \ = 8.75806,
\end{eqnarray}
for the $d$-quark. On the other hand, The sea quark parts are
parameterized as
\begin{equation}
 \Delta_T \bar{q} (x) \ = \ \left[ \,
 a \,e^{-b \,x}  + c \,x^2 \,e^{- d \,x^2} + e \,x^2 + f \,x^3 
 \,\right] \, (1 - x)^g,
\end{equation}
with
\begin{eqnarray}
 a &=& -0.448777, \ \ \ b \ = \ 0.515693, \ \ \ c \ = \ -16.9274, \ \ \ 
 d \ = \ 56.3917, \nonumber \\
 e &=& -14.5186, \ \ \ f \ = \ -5.25201, \ \ \ g \ = 12.2604,
\end{eqnarray}
for the $u$-quark, and with
\begin{eqnarray}
 a &=& 0.439772, \ \ \ b \ = \ 3.0125, \ \ \ c \ = \ 1.28447, \ \ \ 
 d \ = \ 99.8028, \nonumber \\
 e &=& -0.437519, \ \ \ f \ = \ 0.552762, \ \ \ g \ = 2.01257.
\end{eqnarray}
for the $d$-quark.
The 1st moments of these distributions gives the above-mentioned
tensor charges, i.e. $\delta u = 0.95 \,(-0.05)$, $\delta d = - 0.27 \,
(0.08)$, or $g_T^{(I=0)} = 0.68 \,(0.03)$, $g_T^{(I=1)} = 1.21 \,(-0.12)$,
where the numbers in the parentheses are antiquark contributions.
All these distributions should be regarded as initial distributions
given at the low energy scale around $600 \,\mbox{MeV}$.
For obtaining the corresponding transversity distributions at the higher
energy scale, we recommend to use the evolution program at NLO provided
in \cite{HKM98A},\cite{HKM98B} with the starting energy around
$Q^2_{ini} \simeq 0.30 \, \mbox{GeV}^2$.

Now, we show in Fig.1 the CQSM predictions for the transversities and
the longitudinally polarized distributions for the $u$- and $d$-quarks
evolved to the scale $Q^2 \simeq 2.4 \,\mbox{GeV}^2$, which
corresponds to the average energy scale of the global analysis
\cite{ABDKMPT07}.
From this figure, one can clearly see that
the $\Delta_T u (x)$ and $\Delta u (x)$ have nearly
the same magnitude, while the magnitude of $\Delta_T d (x)$ is a factor
of two smaller than that of $\Delta d (x)$.
As already pointed out, this is a reflection of the characteristic
feature $\Delta q^{(I = 0)} (x) \ll \Delta_T q^{(I = 0)} (x)$.

\begin{figure}[htb]
\begin{center}
  \includegraphics[height=.50\textheight]{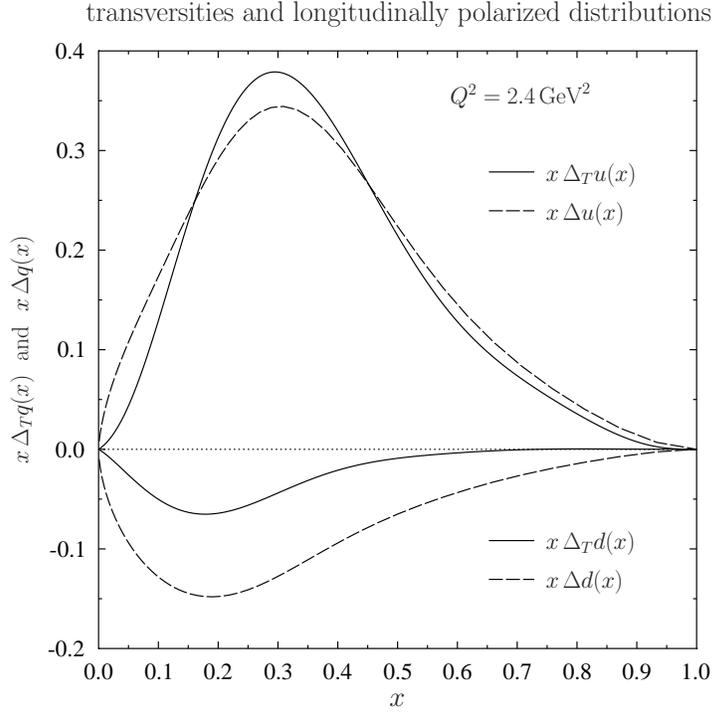}
  \caption{\baselineskip16pt The predictions of the flavor $SU(2)$
  CQSM for the transversities (solid curves) and the longitudinally
  polarized  distribution functions (dashed curves) for the $u$- and
  $d$-quarks evolved to $Q^2 = 2.4 \,\mbox{GeV}$.}%
\label{Fig1}
\end{center}
\end{figure}

Next, let us compare our theoretical predictions for the 
transversities with the global fit by Anselmino {\it et al.}
\cite{ABDKMPT07}.
The two solid curves in Fig.2 stand for the CQSM predictions for the 
transversity distributions $x \Delta_T u (x)$ and $x \Delta_T d (x)$
evolved to $Q^2 = 2.4 \,\mbox{GeV}^2$, while the shaded areas represent
the allowed regions for $x \Delta_T (x)$ and $x \Delta_T d (x)$
in their global fit.
First, one observes that the CQSM prediction for $x \Delta_T d (x)$ is
just within the allowed range of the global fit, whereas the magnitude of 
$x \Delta_T u (x)$ slightly exceeds the upper limit of their fit.
(We shall come back later to this point.)
Next, although the uncertainties of the global fit are still quite large,
a remarkable feature of the transversity distributions seems to be already
seen.

\begin{figure}[bht]
\begin{center}
  \includegraphics[height=.50\textheight]{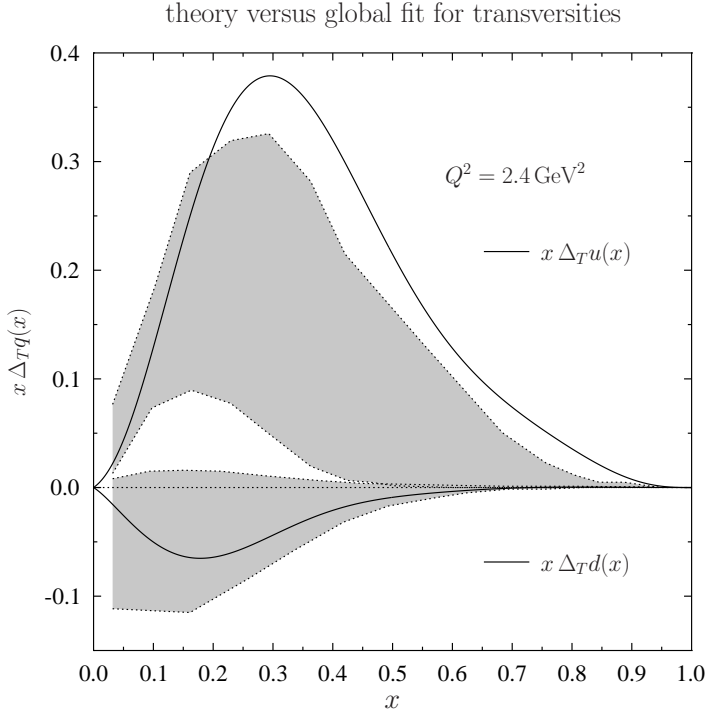}
  \caption{\baselineskip16pt The predictions of the flavor $SU(2)$ CQSM
  for the transversities (solid curves) in comparison with the
  global-fit of \cite{ABDKMPT07} (shaded areas).}%
\label{Fig2}
\end{center}
\end{figure}

The observation that the magnitude of $\Delta_T d (x)$ is much smaller than 
that of $\Delta_T u (x)$ is exactly what the CQSM predicts. As emphasized 
before, the reason can be traced back to the fact that the isoscalar
tensor charge is not so small as the isoscalar axial charge.
Here, one should clearly recognize the following fact. Although almost
all effective models of baryons than the CQSM fail to reproduce
very small axial charge of the order of $0.3 \sim 0.35$, the relatively
large isoscalar tensor charge is a common prediction of many models
including the CQSM. For instance, the MIT bag model (with the constraint
to reproduce $g_A^{(I=1)} = 1.257$) predicts $g_T^{(I=0)} \simeq 0.88$
and $g_T^{(I=1)} \simeq 1.46$ \cite{HJ95}, which turns out to give
remarkably the same numbers as obtained in the relativistic light-cone
quark model \cite{SS97}. The predictions of the hypercentral model
given in \cite{PPB05} are also fairly close the the above
predictions : 
$g_T^{(I=0)} \simeq 0.73$, and $g_T^{(I=1)} \simeq 1.21$.
Also interesting would be the predictions of the lattice
QCD \cite{ADHK97},\cite{Kuramashi98}, which gives
$g_T^{(I=0)} = 0.562 \pm 0.088$ and $g_T^{(I=1)} = 1.07 \pm 0.88$.
We recall that for the axial charges the simulation by the same
group gives $g_A^{(I=0)} = 0.18 \pm 0.10$ and $g_A^{(I=1)} = 0.985
\pm 0.10$, which denotes that $g_A^{(I=0)} \ll g_T^{(I=0)}$,
although the magnitude of $g_A^{(I=1)}$ is obviously underestimated.
Somewhat extraordinary are the predictions of the QCD sum rule
\cite{HJ96}. It predicts $g_T^{(I=0)} = 1.37 \pm 0.55$ and
$g_T^{(I=1)} = 1.29 \pm 0.51$, which dictates that $\delta d$ is
slightly positive. Although this feature itself is not inconsistent
with the result for $\Delta_T d(x)$ obtained in the global fit
\cite{ABDKMPT07}, it would intolerably overestimate the magnitude of
$\Delta_T u(x)$.
In any case, one can now convince that relatively large isoscalar
tensor charge is a common prediction of many effective models.
A uniqueness of the CQSM is that it shares this feature with
these many models, while it is able to reproduce very small
$g_A^{(I=0)}$ or $\Delta \Sigma$. 

The reason why the CQSM predicts very small $g_A^{(I = 0)}$ or 
$\Delta \Sigma$ is very simple. Since it is an effective quark model that 
does not contain the gluonic degrees of freedom explicitly, it satisfies
the nucleon spin sum rule in the following simplified form :
\begin{equation}
 \frac{1}{2} \ = \ \frac{1}{2} \,\Delta \Sigma \ + \ L^Q ,
\end{equation}
with $L^Q$ being the net orbital angular momentum carried by the
quark fields.
On the other hand, according to the physical nucleon picture of
the CQSM as a rotating hedgehog justified in the large $N_c$ QCD,
it predicts very large $L^Q$ around $2 L^Q \simeq 0.65$, which in turn
dictates that $\Delta \Sigma$ is small \cite{WY91}.
As a matter of course, in real QCD, the correct nucleon spin sum
rule contains the gluon contributions as well : 
\begin{equation}
 \frac{1}{2} \ = \ \frac{1}{2} \,\Delta \Sigma \ + \ 
 L^Q \ + \ \Delta g \ + \ L^g .
\end{equation}
However, the recent COMPASS measurement \cite{COMPASS06G} of the
quasi-real photoproduction 
of high-$p_T$ hadron pairs as well as the other independent measurement by 
the PHENIX \cite{PHENIX06} and the STAR collaborations
\cite{STAR05},\cite{STAR06}, 
all indicates that $\Delta g$ is small
at least at the low energy scales of nonperturbative QCD .
Furthermore, the recent NLO QCD analyses by the COMPASS group
as well as the HERMES group with account of the new data on the
spin-dependent structure function of the deuteron indicates
that \cite{COMPASS05D}-\nocite{COMPASS06D}\cite{HERMES06}
\begin{equation}
 \Delta \Sigma \ \simeq \ 0.3 \sim 0.35 ,
\end{equation}
which is now surprisingly close to the theoretical prediction of the
CQSM, as pointed out in \cite{Wakamatsu07}.
Combining all the observations above, one therefore concludes that
the sum of $L_Q$ and $L_g$ must be fairly large at least in the low
energy domain. 

Is there any sum rule which gives a similar constraint on the
magnitude of the isoscalar tensor charge? The answer is partially
yes and partially no. We recall the transverse spin sum
rule (BLT sum rule) proposed by Bakker, Leader and Trueman
\cite{BLT04}, which
in fact contains the transversity distributions as
\begin{equation}
 \frac{1}{2} \ = \ \frac{1}{2} \,\sum_{a = q, \bar{q}} \,
 \int_0^1 \, \Delta_T q^a (x) \ + \ 
 \sum_{a = q, \bar{q}, g} \,\langle L_{s_T} \rangle^a, 
 \label{eq:BLT}
\end{equation}
where $L_{s_T}$ is the component of the orbital angular momentum
$\mbox{\boldmath $L$}$ along the transverse spin direction $s_T$.
Unfortunately, this is not such a sum rule, which is  obtained as
a first moment of some parton distribution functions.
This means that each term of the sum rule does not corresponds
to a nucleon matrix element of a local operator. In fact, 
in the 1st term of the sum rule (\ref{eq:BLT}),
the quarks and antiquark
contributions add, whereas the difference must enter to form the
tensor charge $g_T^{(I = 0)}$. 
In spite of this unlucky circumstance, the theoretical analysis
based on the CQSM strongly indicates that the transversity distributions
for the antiquarks are fairly small, which in turn implies that the
1st term of the sum rule (\ref{eq:BLT}) may not be largely different from
the isoscalar tensor charge $g_T^{(I = 0)}$.
Then, if the feature $g_T^{(I = 0)} \gg g_A^{(I = 0)}$ is in fact 
confirmed experimentally, it would mean that $L_{s_T}^Q + L_{s_T}^g
\ll L^Q + L^g$, i.e, the transverse component of the quark plus
gluon orbital angular momentum is sizably smaller than the
corresponding longitudinal component.
It would certainly provide us with valuable information on the
orbital motion of quarks and gluons inside the nucleon.

At this point, we come back to the observation that the global
fit for $\Delta_T u (x)$ obtained by Anselmino {\it et al.} is
fairly smaller in magnitude than the corresponding prediction of
the CQSM. To get some feeling about the size of the transversities
obtained in their fit, one may attempt to estimate the
tensor charges from their global fit.
Since their fit provides no information on the antiquark
distributions, this is of course possible under the assumption
that the antiquarks contribute little to the tensor charges.
We anticipate that this is not an unreasonable assumption,
since the theoretical analyses based on the CQSM indicates that
the transversity distributions for the antiquarks are fairly
small. Under this assumption, we estimate from the central fit
of \cite{ABDKMPT07} that
\begin{equation}
 \delta u \ \simeq \ 0.39, \ \ \ \ \delta d \ \simeq \ -0.16,
\end{equation}
or equivalently
\begin{equation}
 g_T^{(I = 0)} \ \simeq \ 0.23, \ \ \ \ 
 g_T^{(I = 1)} \ \simeq \ 0.55 ,
\end{equation}
which is understood to hold at $Q^2 \simeq 2.4 \,\mbox{GeV}^2$.
Using the known NLO evolution equation for the first moment of
$\Delta_T q (x)$ \cite{HKK97}-\nocite{KM97}\cite{Vogelsang98},
we can then estimate the tensor charges at
the low energy scale around $Q^2 = 0.30 \,\mbox{GeV}^2 \simeq 
(600 \,\mbox{MeV})^2$.
Here, we use the NLO evolution equation for the 1st moment of 
$\Delta_T q (x)$ given in \cite{HKK97}, which gives
\begin{equation}
 \frac{g_T (Q^2)}{g_T (Q_0^1)}
 \ = \ \left(\,\frac{\alpha (Q^2)}{\alpha (Q_0^2)} 
 \right)^{\frac{\gamma^{(0)}}{2 \,\beta_0}} \,
 \left(\,\frac{\beta_0 + \beta_1 \,\alpha (Q^2)/ 4 \pi}
 {\beta_0 + \beta_1 \,\alpha(Q_0^2)/4 \pi} 
 \right)^{\frac{1}{2} \left(\frac{\gamma^{(1)}}{\beta_2} 
 - \frac{\gamma^{(0)}}{\beta_0} \right)} ,
\end{equation}
where $\alpha (Q^2)$ represents the standard QCD running coupling
constant at the NLO, while
\begin{eqnarray}
 \beta_0 \,\,&=& 11 - \frac{2}{3} \,N_f, \ \ \ 
 \beta_1 = 102 - \frac{38}{3} \,N_f, \\
 \gamma^{(0)} &=& \frac{8}{3}, \hspace{16mm}
 \gamma^{(1)} = \frac{724}{9} - \frac{104}{27} \,N_f,
\end{eqnarray}
with $N_f = 3$.
The result is
\begin{equation}
 \delta u \ \simeq \ 0.49, \ \ \ \ \delta d \ \simeq \ -0.20,
\end{equation}
or 
\begin{equation}
 g_T^{(I = 0)} \ \simeq \ 0.28, \ \ \ \ 
 g_T^{(I = 1)} \ \simeq \ 0.69,
\end{equation}
at $Q^2 = 0.30 \,\mbox{GeV}^2$.
One finds that the magnitudes of $g_T^{(I = 0)}$ and $g_T^{(I = 1)}$
are both roughly a factor of two smaller than the theoretical
predictions of most low energy models as well as
those of the lattice QCD.  What is meant by this discrepancy is not
clear at the moment. Although the global fit carried out in
\cite{ABDKMPT07} is certainly a giant step toward the
experimental extraction of the transversities with minimal
theoretical assumptions, one must certainly be cautious about the
fact that our understanding of the spin-dependent fragmentation
mechanism is still far from complete.
Highly desirable here is some independent experimental information
on the transversity distributions, for instance, from
the Drell-Yan processes \cite{CDL06}.

To sum up, we have carried out a comparative analysis of the
transversities and the longitudinally polarized distribution
functions in light of the new global fit of the transversities and
the Collins fragmentation functions carried out by Anselmino
{\it et al.} \cite{ABDKMPT07}. 
We have pointed out that their result, although with
large uncertainties, already indicates a remarkable qualitative
difference between the transversities and the longitudinally polarized
distributions such that $|\Delta_T d(x) / \Delta d(x)| \ll 
|\Delta d(x) / \Delta u(x)|$, the cause
of which can be traced back to the relation between the isoscalar
axial and tensor charges, $g_A^{(I=0)} \ll g_T^{(I=0)}$.
Combining the standard nucleon spin sum rule and the BLT transverse
spin sum rule \cite{BLT04}, we can further conjecture that the above
relation between the axial and tensor charges would mean
$L^Q_{s_T} + L^g_{s_T} \ll L^Q + L^g$, i.e. the transverse
component of the quark plus gluon orbital angular momentum would be
sizably smaller than the corresponding longitudinal component.
We are not sure yet whether this unique observation can be
understood as a dynamical effect of Lorentz boost
\cite{Melosh74}. Finally, for convenience of future analyses of
DIS processes depending on the transversity distributions, we gave
in the paper the CQSM predictions for the transversities in a simple
parameterized form. They can be used as initial distributions given at
the low energy model scale around $Q^2 \simeq (600 \,\mbox{MeV})^2$.

\vspace{10mm}
\noindent
\begin{large}
{\bf Acknowledgement}
\end{large}

\vspace{3mm}
This work is supported in part by a Grant-in-Aid for Scientific
Research for Ministry of Education, Culture, Sports, Science
and Technology, Japan (No.~C-16540253)

%
%

\setlength{\baselineskip}{5mm}

\end{document}